\documentclass[11pt]{article}
\usepackage{mydef2col}
\usepackage{kantlipsum,widetext}
\usepackage[utf8]{inputenc}
\usepackage[T1]{fontenc}
\usepackage[autostyle=true]{csquotes}
\usepackage{regexpatch}
\usepackage{cprotect}

\usepackage[colorinlistoftodos,prependcaption,textsize=tiny]{todonotes}
\makeatletter
\xpatchcmd{\@todo}{\setkeys{todonotes}{#1}}{\setkeys{todonotes}{inline,#1}}{}{}
\makeatother

\allowdisplaybreaks[4]
\graphicspath{{./Graphs/}}
\usepackage{tikz}
\usetikzlibrary{calc,decorations.markings}
\usetikzlibrary{arrows,patterns,positioning}

\usepackage[authoryear,square]{natbib}
\setcitestyle{square}

\def\calK{{\cal K}}

\def\erf{\operatorname{Erf}}

\def\calG{{\cal G}}

\def\calB{{\cal B}}
\def\calP{{\cal P}}
\def\calL{{\cal L}}

\def\calF{{\cal F}}

\def\LY{ L{\'e}vy }

\newcommand{\Ind}{\mathbf{1}}

\title{Semi-analytic pricing of American options in time-dependent jump-diffusion models with exponential jumps}
\author{\authorstyle{Andrey Itkin}
\newline\newline
\institution{Tandon School of Engineering, New York University, USA}
}

\date{}

\begin{document}

\maketitle

\lettrineabstract{}

\vspace{-1.4in}

%%%%%%%%%%%%%%%%%%%%%%%%%%%%%%%%%%%%%%%%%%%%%%%%%%%%%%%%%%%%
%\section{Introduction}
\vspace{1em}

In an era of machine learning and data science which essentially revolutionized quantitative finance,  one of the crucial and still appealing questions to answer is about how important traditional methods of classical mathematical and computational finance remain to be. As applied to option pricing, a newly prevailing approach deals with first, building a neural network (NN) approximating the "slow" pricer and training it offline for a wide range of the model parameters, and then option prices can be quickly obtained for any set of the parameters by using this NN. In our opinion, indeed computation burden for training the NN is not a problem (since this is done offline), however, the second step seems to be a more delicate issue due to the required output accuracy. The NN is an universal numerical approximator and certainly outputs prices contain various approximation errors. When combined with the errors of the underlying pricer, especially in areas where the solution is notamooth enough, the total level of errors could be vital to make the NN method being not viable. Therefore, we believe that development of accurate analytical methods is an important and still contemporary problem of quantitative finance.

Having said that, in this paper we explore a possibility of constructing semi-analytic solutions for American option prices under jump-diffusion models with exponential jumps. The idea behind this approach is as follows.

Within last few years the author developed and published a series of papers (co-authored with P.~Carr, D.~Muravey and A.~Lipton in various combinations) and also in the book \citep{ItkinLiptonMuraveyBook} known as the Generalized Integral Transforms (GIT) method. This technique originally was used to price various barrier options in a semi-analytic form including double barrier options where the underlying (e.g., the stock price or the interest rate) follow some one-factor model with {\it time-dependent} parameters, or even a stochastic volatility model, see, e.g. \citep{ItkinMuraveySabrJD, CarrItkinMuraveyHeston} and references therein. In all cases, a semi-analytical (or semi-closed form) solution means that first, one needs to solve a linear Volterra integral equation of the second kind to find the gradient of the solution at the moving boundary (or boundaries) as a function of the time. Then the option price is represented explicitly as an integral (could be 1D or 2D) which depends on this gradient.

Moreover, since for some financial models a direct application of these methods does not provide promising results (for instance, this is the case for popular among practitioners the Black-Karasinski model), an extension of the GIT method called a Multilayer GIT method for solving one-factor parabolic equations was developed in  \citep{ItkinLiptonMuraveyMulti,ML2} which presents a powerful alternative to the well-known finite difference and Monte Carlo methods, and combines semi-analytical and numerical techniques to provide a fast and accurate way of finding solutions to the corresponding equations.

The second piece of the puzzle lies in the fact that as shown in \citep{CarrItkin2020jd}, the same technique can be applied to pricing American options. Then in a recent paper \citep{ItkinMuravey2023} we have shown in detail how the GIT method can be used to obtain semi-analytical expressions for the American option prices in various {\it time-dependent one-factor diffusion} models.

A natural question to ask would be whether this technique can also be used for jump-diffusion models. Therefore, in this paper we are trying to answer it (positively). But to complete the puzzle of pricing American options under time-dependent jump-diffusion models in a semi-analytical form, we need another piece which is represented in detail in our book \citep{ItkinBook}, but first was introduced in \citep{ItkinCarr2012Kinky}. The main idea is to reduce a pricing partial integro-differential equation (PIDE) (which naturally appears when pricing options under jump-diffusion models) to a {\it pseudo-parabolic} partial differential equation (PDE).

In more detail, as per a foreword of Peter Carr to \citep{ItkinBook}, "the assumption that the underlying price process is a jump diffusion caused option pricing to reduce to the solution of either a PIDE or a partial differential difference equation (PDDE). In both cases, the operator acting on the option price ceased to be local. It, therefore, became necessary to know the numerical value of option prices away from a particular space-time point in order to propagate option prices at that point. This non-locality slowed down the numerical evaluation of option prices, discouraging the adoption of realistic jump models  in high performance environments such as automated option market making. And it was universally  accepted that the adoption of jumps to enhance financial realism carried with it the greater computational burden associated with non-local valuation operators". But in \citep{ItkinCarr2012Kinky} it was discovered "... that there exist  special types of jump processes for which the advantages of a local operator need not be foregone.  For certain kinds of realistic jump processes detailed in \citep{ItkinBook},  one can use either the usual non-local valuation operator or an unusual local valuation operator.  When the choice between the two operators  is available, the local valuation operator will always be the more computationally efficient choice".

To summarize what has been already discussed in above, our plan to attack the problem entitled this paper is as follows:
\begin{enumerate}
\item  pick a simple jump-diffusion model - exponential jumps, but assume all the model parameters to be time-dependent;

\item consider an American option written on the underlying following this jump-diffusion model.

\item write the corresponding pricing PIDE valid in the continuation (holding) region in the form of a pseudo-parabolic PDE using a local pseudo-differential operator, \citep{ItkinBook};

\item transform this pseudo-parabolic PDE to another non-parabolic PDE, e.g. those used to describe ﬁltration of a compressible ﬂuid in a cracked porous medium or unsteady motions of second-grade non-Newtonian ﬂuids, see \citep{Polyanin2002} and references therein;

\item construct an appropriate GIT to derive a Volterra-type integral equation(s) for the exercise boundary $y(\tau)$ which is a function of the backward time $\tau$ only, \citep{ItkinMuravey2023};

\item discuss an efficient numerical method to solve this integral equation(s).

\end{enumerate}

This machinery could be more efficient numerically than the standard methods (i.e., finite-difference methods, \citep{ItkinBook}), and provide higher accuracy, with an extra power when computing option's Greeks, \citep{ItkinLiptonMuraveyBook}.

It  is worth mentioning that our approach is close in spirit to that in \citep{Chiarella2009} (further CZ). The differences are as follows. First, we use time-dependent parameters of the model while in CZ they are constant. Accordingly, the Fourier transform method utilized in CZ is not applicable here.  Second, in CZ a Merton jump-diffusion model is used while here we consider exponential and double exponential jumps. Third, our method allows finding the exercise boundary $x_B(t)$ as a function of the time $t$ by using a sequential in time procedure where at every time step $x_B(t)$ and $P_{xx}(t, x_B(t))$ solve a system of two equations. It can be solved iteratively: given an initial guess for the option Gamma at the boundary $P_{xx}(t, x_B(t))$, $x_B(t)$ solves an algebraic equation. The solution can be substituted into the other (integral) equation for the option Gamma which then becomes a {\it linear} Fredholm-Volterra equation of the second kind and can be solved by using, e.g., some kind of collocation methods. This gives the next approximation to the Option Gamma, so the whole loop can be continued until converges. Then the Put option price is given analytically in closed form. In contrast, in CZ the authors have to solve a system of two nonlinear integral equations which is much more expensive.

\section{Pricing PIDE}

Let us consider time evolution of a stock with the price $S_t$ where $t$ is the time. Assume that $S_t$ follows a simple jump-diffusion model
\begin{align} \label{sde}
d S_t &= [r(t)  - q(t)] S_t dt + \sigma(t) S_t W_t + S_t d L_t, \qquad S_0 \equiv S, \, (t, S) \in [0,\infty) \times [0,\infty).
\end{align}
Here $r(t), q(t), \sigma(t)$ are the deterministic short interest rate, dividend yield and volatility, $W_t$ is the Brownian motion. The process $L_t$ is a pure discontinuous jump process with generator $A$, and a \LY measure $\nu (dy)$
\begin{equation}
A f(x) = \int_{\mathbb{R}}\left( f(x+y) - f(x) - y \mathbf{1}_{|y|<1}\right) \nu (dy), \qquad
\int_{|y|>1}e^{y}\nu (dy)<\infty,
\end{equation}
\noindent where $x \equiv \log (S_t/S_*)$, $S_*$ - some normalization constant.

It is known that the price of a European (e.g., Put) option $P(t, x)$, written on the underlying $S_t$, solves the PIDE, \citep{ContTankov}
\begin{align} \label{PIDE}
r P(t,x) = \fp{P(t,x)}{t} &+ \left[r(t) - q(t) - \frac{1}{2}\sigma^2(t) \right] \fp{P(t,x)}{x} + \frac{1}{2}\sigma^2(t) \sop{P(t,x)}{x} \\
&+  \int_\mathbb{R}\left[ P(t,x+y) - P(t,x) - (e^y-1)\fp{P(t,x)}{x} \right] \nu(dy), \qquad
(t, x) \in [0,T] \times \mathbb{R}, \nonumber
\end{align}
The jump integral in the RHS of \eqref{PIDE} is a non-local operator so computing it numerically is challenged and computationally expensive. This PIDE has to be solved subject to the terminal condition $P(T, x) = H(x)$, where $H(x)$ is the option payoff at maturity $T$, and some boundary conditions which depend on the type of the option. The solution belongs to the class of viscosity solutions \citep{ContTankov}.

To solve \eqref{PIDE} various numerical method can be used, e.g., a finite-difference (FD) method, as this is described in detail in \citep{ItkinBook} (see also references therein). However, for the American options the method has to be equipped with additional checks for the optimality of the option exercise at every given stock level. Thus, the exercise boundary $S_B(t)$ which is not known in advance, has to be found simultaneously with the option price as a part of the solution, again see \citep{ItkinBook} and references therein.

We now transform \eqref{PIDE} to the form of a local pseudo-parabolic PDE by using the method of \citep{ItkinBook}. It is shown there that the integral term in \eqref{PIDE} can be represented by using a well-known fact from quantum mechanics, \citep{OMQM}, namely: that a translation (shift) operator in $L_2$ space can be expressed as follows
\begin{equation} \label{transform}
\mathcal{T}_b = \exp \left( b \dfrac{\partial}{\partial x} \right), \qquad
\mathcal{T}_b f(x) = f(x+b),
\end{equation}
\noindent with $b$ = const. This can be easily proved by using a Taylor series expansion of both parts of \eqref{transform}.

\begin{figure}[!htb]
\begin{center}
\subfloat[]{\includegraphics[width=0.53\textwidth]{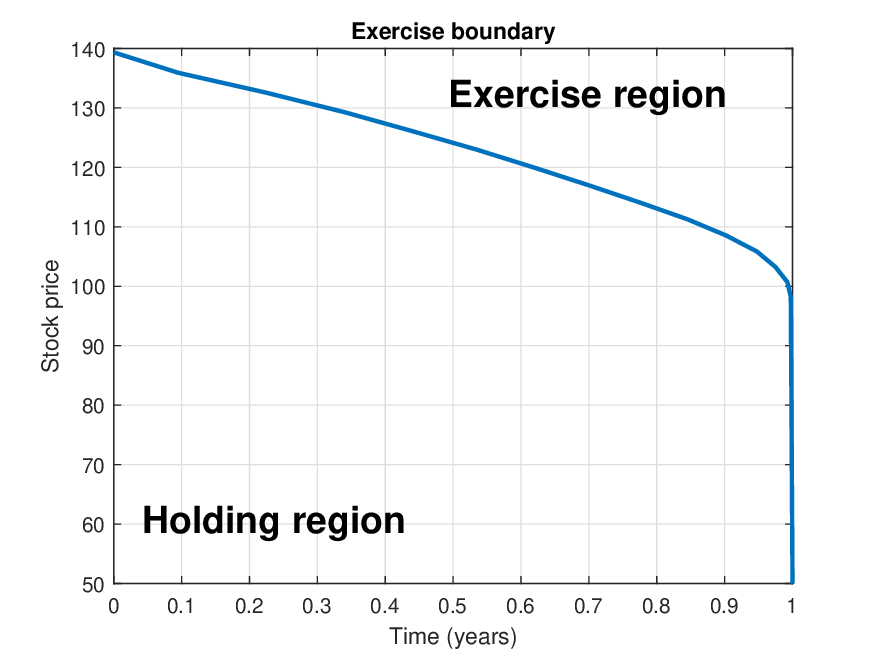}}
\hspace*{-0.3in}
\subfloat[]{\includegraphics[width=0.53\textwidth]{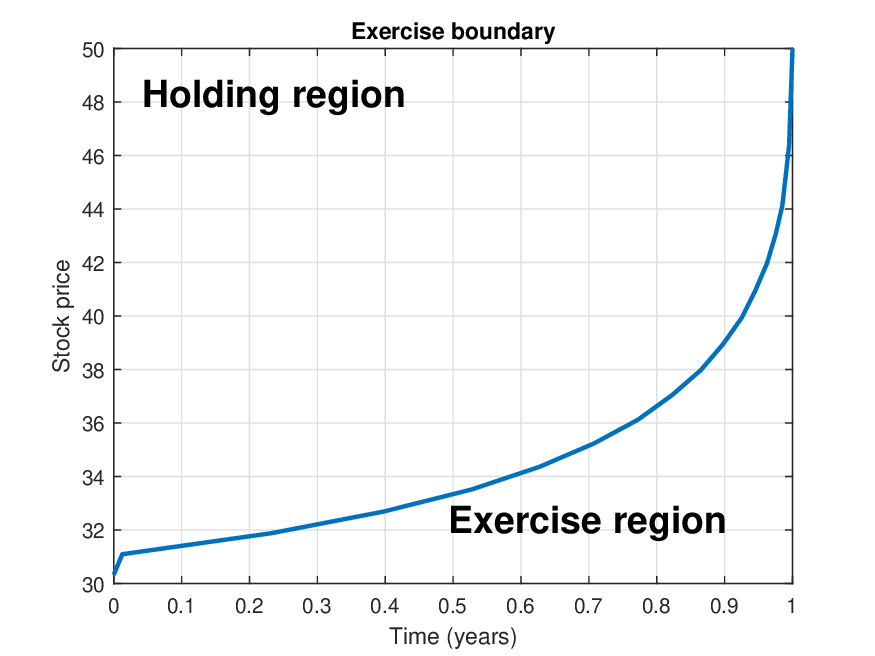}}
\end{center}
\vspace{-1em}
\caption{Typical exercise boundary for the American Call option (a) and Put option (b) under the Black-Scholes model with $K = 50, r = 0.2, q = 0.1, \sigma = 0.5, T = 1$. }
\label{EB}
\end{figure}

To proceed, let us notice that there exists an alternative approach to finding prices of the American options which utilizes a notion of the exercise boundary, \citep{ItkinMuravey2023}. By definition, the boundary $S_B(t)$ is defined in such a way, that, e.g., for the American Put option $P(S,t)$ at $S \le S_B(t)$ it is always optimal to exercise the option, therefore  $P(S,t) = K - S$. For the complementary domain $S > S_B(t)$ the earlier exercise is not optimal, and in this domain $P(S,t)$ solves the corresponding Kolmogorov equation. This domain is called the continuation (holding) region. For instance, for the American Put we have $P(S_B(t),t)  = K - S_B(t)$, and for the American Call - $C(S_B(t),t)  = S_B(t) - K$. A typical shape of the exercise boundary for the Put option is presented in Fig.~\ref{EB}.
Thus, in the continuation region which is bounded by the exercise boundary $S_B(t)$, the American option price still solves \eqref{PIDE}.

Thus, in what follows we consider \eqref{PIDE} only in the continuation region. Accordingly, in this region all derivatives $P^{(n)}_x(x), \ n > 0$ in \eqref{transform} are well-defined.

Therefore, the integral in Eq.~(\ref{PIDE}) can be formally rewritten as
\begin{align} \label{intGen}
\int_\mathbb{R} \left[ P(t,x+y) \right. & \left. - P(t,x) - (e^y-1) \fp{P(t,x)}{x} \right] \nu(dy) =  \mathcal{J} P(t,x), \\
\mathcal{J} & \equiv \int_\mathbb{R}\left[
\exp \left( y \dfrac{\partial}{\partial x} \right) - 1 - (e^y-1) \fp{}{x} \right] \nu(dy). \nonumber
\end{align}

In the definition of operator $\mathcal{J}$ (which is actually an infinitesimal generator of the jump process), the integral can be formally computed under some mild assumptions about existence and convergence if one treats the term $\partial/ \partial x$ as a constant. Therefore, the operator $\mathcal{J}$ can be considered as some generalized function of the differential operator $\partial_x$. We can also treat $\mathcal{J}$ as a pseudo-differential operator. An important point, however, is that $\mathcal{J}$ is a local operator in the spatial variable's space\footnote{The jump operator in \eqref{intGen} is working as kind of an expectation. Therefore, despite a bit surprising, that expectation over $y$ of a non-local operator in variable $y$ could be represented as some local, but now pseudo-differential, operator (in more detail, see \citep{ItkinBook}).}.

With allowance for this representation, the whole PIDE in the \eqref{PIDE} can be re-written in the operator form as, \citep{ItkinCarr2012Kinky}
\begin{equation} \label{oper}
\partial_t P(t,x) = [\mathcal{D} + \mathcal{J}]P(t,x),
\end{equation}
\noindent where the operator $\mathcal{D}$ is an infinitesimal generator of diffusion. Thus, by applying \eqref{transform} to the jump integral we transform the original linear PIDE (which is {\it non-local} because the jump (integral) operator is non-local) into a linear pseudo-differential equation where the RHS operator in \eqref{oper} is {local but now a pseudo-differential operator}.

Notice that for jumps with finite variation and finite activity, the last two terms in the definition of the jump integral $\mathcal{J}$ in \eqref{PIDE} could be integrated out and added to the definition of $\mathcal{D}$. In the case of jumps with finite variation and infinite activity, the last term could be integrated out. However, here we will leave these terms under the integral because this transformation (moving some terms under the integral to the diffusion operator) does not affect our method of computing the integral.

\section{Exponential jumps} \label{ExpJumps}

First, let us consider only negative exponentially distributed jumps\footnote{For the positive jumps this could be done in a similar way. The denominator in \eqref{expJI} then changes to $\phi-1$ and the term $\phi + \triangledown_x$ changes to $\phi  - \triangledown_x$ with $\phi > 1$.}, see \citep{Lipton2002a} among others, i.e.
\begin{equation} \label{expJ}
\nu(J) =
\begin{cases}
\phi e^{\phi J}, & J \le 0 \cr
0, & J > 0,
\end{cases}
\end{equation}
\noindent where $\phi > 0$ is the parameter of the exponential distribution. With the \LY measure $\nu(dy)$ given in \eqref{expJ} and the intensity of jumps (of the Poison process) $\lambda \ge 0$  we can substitute $\nu(dy)$ into \eqref{intGen} and integrate. The result reads
\begin{equation} \label{expJI}
\mathcal{J} = \dfrac{\lambda}{\phi+1}(\phi + \triangledown_x)^{-1} (\triangledown_x^2 - \triangledown_x), \qquad \triangledown_x \equiv \partial_x.
\end{equation}

Let us consider an American Put option written on the underlying $S_t$ evolving by \eqref{sde} with the strike price $K$. To find the Put option price $P(t,x)$ one has to solve \eqref{PIDE} together with \eqref{intGen} and \eqref{expJI} to obtain
\begin{align} \label{PIDE1}
\left( \fp{}{t}  + \calL \right) P(t,x) &= 0, \qquad (t,x) \in \Omega_x: [0,T] \times [x^+_B(t), \infty),
\qquad x_B(t) \equiv \log \frac{S_B(t)}{S_*}, \\
\calL &= -[r(t) + \lambda] + \left[r(t) - q(t) - \frac{1}{2}\sigma^2(t) + \frac{\lambda}{1 + \phi}\right] \triangledown_x + \frac{1}{2}\sigma^2(t) \triangledown^2_x + \lambda \phi (\phi + \triangledown_x)^{-1}. \nonumber
\end{align}
\noindent where $\triangledown_x \equiv \partial_x$ and $\calL$ is a pseudo-differential operator since its last term is proportional to a resolvent $(\phi + \triangledown_x)^{-1}$ \footnote{Since the exponential jumps belong to a class of jumps with finite variation and finite activity, integral of each term in \eqref{intGen} exists and can be computed separately.}.

In what follows, we extend this model by assuming the jump parameters $\phi, \lambda$ to be time-dependent for two reasons. First, there exists vast literature that, by analyzing the market data, finds strong empirical support for time-varying jump intensities, see \citep{Cirano2009, Hanson2003} among others. Second, as it can be seen below, the assumption $\phi = \phi(t), \ \lambda = \lambda(t)$ doesn't affect tractability of our approach.

Assuming the resolvent exists, is analytic and bounded, after some algebra we can rewrite \eqref{PIDE1} in the form
\begin{align} \label{PDE}
0 &= \fp{u}{t}  - a_d(t) \fp{u}{x} + a_v(t) \sop{u}{x} - a_s(t)  u  + a_j(t) P(t,x), \\
u(t,x) &\equiv \phi(t) P(t,x) + P_x(t,x), \qquad (t,x) \in \Omega_x, \nonumber \\
a_d(t) &= r(t) - q(t) - \frac{1}{2}\sigma^2(t) + \frac{\lambda(t)}{1 + \phi(t)}, \qquad
a_v(t) = \frac{1}{2}\sigma^2(t),  \nonumber \\
a_s(t) &= r(t) + \lambda(t), \qquad
a_j(t) = \lambda(t) \phi(t) - \phi'(t). \nonumber
\end{align}

It can be seen that out trick reduces the {\it PIDE} in \eqref{PIDE} to the {\it PDE} in \eqref{PDE} which, however, is not anymore, a parabolic PDE  with respect to  $P(t,x)$ (which is a traditional form of PDEs in mathematical finance). Equations of the type \eqref{PDE} have been shown to have application in the unidirectional flow of a second-grade fluid, statistical mechanics and the study of heat conduction, see \citep{Momoniat2015, Polyanin2002} and references therein. Also, this type of PDEs can be associated with the integrable Camassa--Holm equation used to model the solution wave interaction together with the breaking of wave, \citep{Yu2015}. Pseudo-parabolic equations can also be regarded as a Sobolev-type equation or a Sobolev–Galpern-type equation, see discussion and other applications in \citep{Zhou2021}.

The \eqref{PDE} holds in the continuation region $\Omega_x$ and has to be solved subject to the terminal condition at the option maturity $t=T$ \footnote{It is worth noticing that the value $P(T^-,x)$ might differ from $P(T,x)$, i.e. the exercise boundary $S_B(t)$ may have a jump at $t=T$, \citep{Kwok2022,Chiarella2009}.}
\begin{equation} \label{tc0}
P(T,x) = S_*(e^{k} - e^{x})^+,
\end{equation}
\noindent where $k = \log (K/S_*)$ is the log-strike, and the boundary conditions. For the American Put the boundary conditions read, \citep{Kwok2022, ItkinMuravey2023}
\begin{equation} \label{bc0}
P(t,x)\Big|_{x \uparrow \infty} = 0, \qquad P(t,x_B(t)) = S_*\left( e^{k} - e^{x_B(t)}\right).
\end{equation}

Since $(t, x) \in \Omega_x$ (this is the holding area under the exercise boundary curve in Fig.~\ref{EB} in the original coordinates $(t, S)$ ), and $S_B(T) = K$,  the terminal condition in \eqref{tc0} becomes homogeneous, i.e.
\begin{equation} \label{tcFin}
P(T,x) = 0.
\end{equation}
Also, it follows from \eqref{bc0} that
\begin{equation} \label{bc1}
P_x(t,x_B(t)) = - S_* e^{x_B(t)}.
\end{equation}

With arbitrary time-dependent coefficients it is hard to solve \eqref{PDE} analytically. However, since our main goal is to determine a shape of the exercise boundary, we proceed with a trick. The trick is: first, to consider a "homogeneous" version of \eqref{PDE} with respect to $u(t,x)$ only, and treat the last terms in the RHS as a source term. After the solution of the homogeneous PDE is found, the source term can be taken into account by using the Duhamel's principle for PDEs with moving boundaries, \citep{ItkinMuravey2023}.

\subsection{Solution for the American Put} \label{AmerOptSec}

\subsubsection{The homogeneous PDE} \label{secHomog}

Here we consider a homogeneous version of \eqref{PDE} which reads
\begin{align} \label{PDE1}
0 &= \fp{u}{t}  - a_d(t) \fp{u}{x} + a_v(t) \sop{u}{x} - a_s(t)  u.
\end{align}
Since \eqref{PDE1} is written in terms of the function $u(t,x)$, the terminal and boundary conditions for $u(t,x)$ can be obtained from \eqref{tcFin}, \eqref{bc0} to yield
\begin{align} \label{bcU}
u(T,x) = - S_* \gamma(x) e^{x}, \quad u(t,x)\Big|_{x \uparrow \infty} = 0, \quad u(t,x_B(t)) = S_* \left[ \phi(t) e^k -  (\gamma(x) +\phi(t)) e^{x_B(t)} \right],
\end{align}
\noindent where $\gamma(x) = \Ind_{x=k}/2$. As per \citep{Polyanin2002}, the transformation
\begin{align} \label{trans}
u(t, x) &= U(\tau, z) h(t), \hspace{0.25em} \tau =  \int_t^T  a_v(s) ds, \hspace{0.25em} z = x + f(t), \hspace{0.25em}
h(t) = \exp \left[ \int_T^t  a_s(k) dk \right], \hspace{0.25em} f(t) = \int_T^t a_d(s) ds,
\end{align}
\noindent converts \eqref{PDE1} to a heat equation problem
\begin{align} \label{Heat}
U_\tau &= U_{zz},  \qquad U(0,z) = -\frac{1}{2} S_*e^{k}\Ind_{z=k+f(\tau)}, \qquad U(\tau,z)\Big|_{z \uparrow \infty} = 0, \qquad U(\tau, y(\tau)) = g(\tau),  \\
y(\tau) &\equiv z_B(\tau) = x_B(\tau) + f(\tau), \qquad
g(\tau) = \frac{S_*}{h(t(\tau))} \left\{\phi(t(\tau)) e^k  -  \left[\gamma(y(\tau)-f(\tau))+\phi(t(\tau))\right] e^{y(\tau)- f(\tau)} \right\}. \nonumber
\end{align}

Fortunately, this problem has been already solved in \citep{ItkinMuravey2023}. The solution reads
\begin{align} \label{uFourier}
U(\tau, z) &\equiv G(\tau, z, y(\tau)) = - S_* \frac{e^{k}}{4\sqrt{\pi \tau}} \int_{y(0)}^{\infty} \Ind_{\xi=k+f(\tau)} \left[e^{-\frac{( \xi - z)^2}{4\tau}} -  e^{-\frac{(\xi + z -2 y(\tau))^2}{4\tau}} \right] d\xi  \\
&-\int_0^\tau \frac{\Psi(s,y(s)) + y'(s) g(s)}{2\sqrt{\pi (\tau - s)}} \left[e^{-\frac{(z - y(s))^2}{4(\tau-s)}} -  e^{-\frac{(z -2 y(\tau) + y(s))^2}{4(\tau-s)}} \right] ds  \nonumber \\
&+ \int_0^\tau \frac{g(s)}{4\sqrt{\pi(\tau - s)^3}} \left[ (z - y(s)) e^{-\frac{(z - y(s))^2}{4(\tau -s)}} +
(y(s) + z - 2 y(\tau)) e^{-\frac{(z + y(s) - 2 y(\tau))^2}{4(\tau -s)}} \right] ds. \nonumber
\end{align}
Since $y(0) = k, f(0) = 0$, the first term (integral) in the RHS can be represented as follows
\begin{align*}
- S_* \frac{e^{- f(\tau)}}{2\sqrt{\pi \tau}}   \int_{y(0)}^{\infty} \gamma(\xi - f(\tau)) e^{\xi} \left[e^{-\frac{( \xi - z)^2}{4\tau}} -  e^{-\frac{(\xi + z -2 y(\tau))^2}{4\tau}} \right] d\xi &=
- S_* \frac{e^{k-f(\tau)}}{4\sqrt{\pi \tau}}
\left[e^{-\frac{( k - z)^2}{4\tau}} -  e^{-\frac{(k + z -2 y(\tau))^2}{4\tau}} \right].
\end{align*}

In \eqref{uFourier} $\Psi(\tau, y(\tau))$ is the gradient of the solution at the moving boundary $y(\tau)$
\begin{equation} \label{psiDef}
\Psi(\tau, y(\tau)) =  \fp{U(\tau, z)}{z} \Big |_{z = y(\tau)}.
\end{equation}

Using the definitions in \eqref{PDE}, \eqref{trans}, one can find this gradient in an explicit form
\begin{align*}
\fp{U(\tau, z)}{z} &= \frac{1}{h(\tau)} \left[ \phi(\tau) P_z(\tau, z)  + P_{zz}(\tau, z) \right].
\end{align*}
By the smooth fit principle, \citep{Kwok2022} (also, see a survey in \citep{Chiarella2009}) the option value and the first space derivative are continuous at the exercise boundary, while the second derivative experiences a jump. Hence, the value of $P_{zz}(\tau, y(\tau))$ has to be understood here as $P_{zz}(\tau, z \to y(\tau)^+)$, i.e. as the limit of $P_{zz}(\tau, z)$ in the continuation region at $z \to y(\tau)$.

Since $P_z(\tau, z)$ is known, we can re-write \eqref{psiDef} in the form
\begin{equation} \label{Psi}
\Psi(\tau, y(\tau)) = \frac{1}{h(\tau)}\left[P_{zz}(\tau, z)\Big |_{z = y(\tau)^+} - S_* \phi(\tau) e^{y(\tau) - f(\tau)}\right].
\end{equation}

Substituting the definition of $U(\tau,z)$ into \eqref{uFourier}, we obtain an ordinary differential equation (ODE) for $P(\tau, x)$
\begin{align} \label{ode}
\phi(\tau) P(\tau, x) + P_x(\tau, x) &= h(\tau) G(\tau, x + f(\tau), x_B(\tau) + f(\tau)),
\end{align}
\noindent and then integrating in $x$, find
\begin{align} \label{solHom}
P(t,x) &= e^{- \phi(t) x} \left\{ S_*\left( e^{k} - e^{x_B(t)} \right) e^{\phi(t) x_B(t)} + h(t) \int_{x_B(t)}^x e^{\phi(t) \eta} G(\tau(t), z(\eta), x_B(t)) d \eta \right\}.
\end{align}
To underline, this is not a closed form solution for $P(t,x)$ but rather an equation since $G(\tau, z, y(\tau))$ under the integral in the RHS is a function of $P_{zz}(\tau, y(\tau))$.

To proceed, we differentiate twice both sides of this equation by $x$, and substitute $x \to x_B(t)$ to obtain
\begin{align} \label{solHom2}
P_{xx}(t,x_B(t)) &= S_* \phi^2(t)\left( e^{k} - e^{x_B(t)} \right) - h(t) \Big[ \phi(t) G(\tau(t), z(x_B(t)), x_B(t)) - G_z(\tau(t), z(x_B(t)), x_B(t)) \Big].
\end{align}
After some algebra, this equation can be re-written in the form, see Appendix~\ref{app3}
\begin{align} \label{PxxVolterraFin}
P_{xx}&(\tau,x_B(\tau)) = \calF(\tau,x_B(\tau)) - \int_0^\tau P_{xx}(s, y(s)) \frac{h(\tau) (y(\tau) -y(s)) }{2 h(s)} \frac{e^{-\frac{(y(\tau) - y(s))^2}{4(\tau-s)}}}{\sqrt{\pi (\tau - s)^3}} ds, \\
\calF&(\tau,x_B(\tau)) = S_* \phi^2(\tau)\left( e^{k} - e^{x_B(\tau)} \right) - h(\tau) \Bigg\{ g(\tau) +
S_* \frac{e^{- f(\tau)}}{2\sqrt{\pi \tau^3}} \int_{y(0)}^{\infty} \gamma(\xi) e^{\xi}   (y(\tau) -\xi) e^{-\frac{(y(\tau)-\xi)^2}{4\tau}}d\xi \nonumber \\
&- \int_0^\tau \frac{e^{-\frac{(y(\tau) - y(s))^2}{4(\tau-s)}}}{2\sqrt{\pi (\tau - s)^3}}
\Bigg[ \frac{S_*}{h(s)} \phi(s)  (y(\tau) -y(s)) e^{y(s) - f(s)} + g(s) \left[ 1 - \frac{(y(\tau) - y(s))^2}{2 (\tau - s)} + y'(\tau) \right] \Bigg]
ds \Bigg\}. \nonumber
\end{align}

\paragraph{The boundary $x_B(t)$ is known.}  If the boundary $x_B(t)$ is known, \eqref{PxxVolterraFin} becomes a linear Volterra equation of the second type for $P_{xx}(t, x_B(t))$. Thus, the option Gamma $P_{xx}(\tau,x_B(\tau))$ can be found by solving \eqref{solHom2}. This can be done by using any numerical method suitable to solving linear integral Volterra equations of the second type, (see a short survey in \citep{ItkinMuravey2023}). Then, substituting the expressions for $x_B(t)$ and $P_{xx}(t,y(\tau))$ into the definition of $\Psi(\tau, y(\tau))$ in \eqref{Psi}, and then into \eqref{uFourier}, we obtain an explicit representation for the function $G(\tau, z, y(\tau))$. Then the value of the American Put option price $P(t,x)$ follows from \eqref{solHom}.

Also, based on the definition on $G(\tau(t), z(\eta), x_B(t))$ in \eqref{uFourier}, the last integral in \eqref{solHom} on $\eta$ can be taken in closed form if $P_xx(t, x_B(t)$ is known. This is shown in Appendix~\ref{app2}.

\paragraph{The boundary $x_B(t)$ is not known.} This is the usual case for the American option.

Since $x_B(t)$ is not known, \eqref{PxxVolterraFin} should be supplemented by an extra one, so solving it together would provide both $x_B(t)$ and $P_{xx}(t,x_B(t))$. One can think about using \eqref{solHom} by first, differentiating both parts by $x$, and then setting $x \to x_B(t)$ (since $P_x(t,x_B(t))$ is known in closed form). This is the standard way used when constructing semi-analytical solution for barrier options, \citep{ItkinLiptonMuraveyBook}. However, here instead of an equation for $x_B(t)$ this approach gives rise to a valid identity due to the boundary condition for $u(t,x)$ at $x = x)B(t)$ in \eqref{bcU}. Therefore, the homogeneous problem for American options cannot be solved in such a way.

This can be understood by an observation that the real problem in \eqref{PIDE1} is inhomogeneous unless $\lambda(t) = 0$. However, at $\lambda(t) = 0$ the last term in the definition of $\calL$ in \eqref{PIDE1} disappears. Therefore, the transformation from \eqref{PIDE1} to \eqref{PDE} cannot be done, and the definition of $u(t,x)$ in \eqref{PDE} should be changed. Thus, it is expected that the solution of the inhomogeneous problem won't be correct in the singular limit $\lambda = 0$.

\subsubsection{The inhomogeneous PDE in \protect\eqref{PDE}}

For solving a non-homogeneous equation in \eqref{PDE} let us use the same transformations as in \eqref{trans}. Then, instead of the problem in \eqref{Heat} we obtain a similar problem
\begin{align} \label{HeatNonHom}
U_\tau &= U_{zz} + \Lambda(\tau, z), \qquad
\Lambda(\tau, z) = \frac{a_j(\tau)}{h(\tau)} P(\tau,z).
\end{align}

By using the Duhamel's principle for PDEs with moving boundaries, \citep{ItkinMuravey2023} (also, see Appendix~\ref{secDuhamel}), it can be shown that the source term $\Lambda(\tau, z-f(\tau))$ in \eqref{PDE} translates into an extra term $J(\tau,z)$ in the definition of $G(\tau, z, y(\tau))$ in \eqref{uFourier}, \citep{ItkinMuravey2023}
\begin{align}
J(\tau,z) &= \int_0^\tau \int_{y(s)}^{\infty}\frac{a_j(s) }{h(s)}P(s,\xi) \calG(\xi,s, z, \tau) d\xi ds, \\
 \calG(\xi,s, z, \tau) &= \frac{1}{2 \sqrt{\pi (\tau - s)}} \left[e^{-\frac{(\xi - z)^2}{4(\tau-s)}} -  e^{-\frac{(\xi -2 y(\tau) + z)^2}{4(\tau-s)}} \right]. \nonumber
\end{align}
By substituting this updated definition of $G(\tau, z, y(\tau))$ into \eqref{solHom}, we obtain
\begin{align} \label{solInhom}
P&(t,x) = e^{- \phi(t) x} \Bigg\{ S_*\left( e^{k} - e^{x_B(t)} \right) e^{\phi(t) x_B(t)} + h(t) \int_{x_B(t)}^x e^{\phi(t) \eta} G(\tau(t), z(\eta), x_B(t)) d \eta \\
&+ h(t) \int_0^\tau \int_{x_B(s)}^{\infty} \frac{a_j(s) P(s,\xi)}{h(s)} \calK_1(\tau, x, s, \xi) d\xi ds  \Bigg\}, \quad
\calK_1(\tau, x, s, \xi) \equiv \int_{x_B(\tau)}^{x} e^{\phi(\tau) \eta} \calG(\xi,s, z(\eta), \tau) d\eta,  \nonumber
\end{align}
\noindent where $G(\tau(t), z(\eta), x_B(t))$ is defined in \eqref{uFourier}. The integral in the definition of $\calK_1(\tau, x, s, \xi)$ can be taken in closed form, see Appendix~\ref{app2}.

The RHS of \eqref{solInhom} depends on $y(\tau)$, $P_{zz}(\tau, y(\tau))$ and also $P(s, \xi)$ in the RHS under the integral. Therefore, if $x_B(t)$ is known, \eqref{solInhom} is an integral Volterra equation of the second kind with respect to $P(t,x)$. Moreover, $G(\tau, z, y(\tau))$ in this equation depends on $P_{xx}(t, x_B(t))$.

One can also twice differentiate both parts of \eqref{solInhom} by $x$ and set $x = x_B(\tau)$ to obtain
\begin{align} \label{solInhomSup1}
P_{xx}(\tau, x_B(\tau)) = - \phi(\tau) P(\tau, x_B(\tau)) +  h(\tau) G_x(\tau, z(x_B(\tau)), x_B(\tau)) +  J_x(\tau,x)]\Big|_{x=x_B(\tau)}.
 \end{align}
It turns out that given $x_B(t)$, \eqref{solInhomSup1} is actually a {\it linear} integral Volterra equation of the second kind for $P_{xx}(t, x_B(t))$. Indeed, $P(\tau, x_B(\tau))$ is known and given in \eqref{bc0}, and by the definition of $\calG(\xi, s, z, \tau)$ we have
\begin{align*}
\calG(\xi, s, x_B(t)+f(\tau), \tau) &=
\begin{cases}
0, & s \ne \tau, \\
\delta(\xi - x_B(t) - f(\tau)), & s = \tau.
\end{cases}
\end{align*}
Therefore, the last double integral in \eqref{solInhomSup1} transforms to
\begin{align} \label{intJ}
J(\tau,x) &= \int_0^\tau \int_{x_B(s) + f(s)}^{\infty} \frac{a_j(s)}{h(s)} P(s,\xi) \calG(\xi,s,x+f(\tau),\tau) d\xi ds = \frac{a_j(\tau)}{h(\tau)} S_* \left(e^k - e^{x}\right).
\end{align}
Accordingly, $J_x(\tau,x_B(\tau)) = -\frac{a_j(\tau)}{h(\tau)} S_* e^{x_B(\tau)}$.

If $x_B(t)$ is not known, \eqref{solInhomSup1} can be supplemented by another equation obtained from \eqref{solInhom} by differentiating both parts by $x$ and setting $x = x_B(t)$. This yields
\begin{align} \label{solInhomSup}
P_x(t,x_B(t)) + \phi(t) P(t, x_B(t)) =  h(t) \left[ G(\tau(t), z(x_B(t)), x_B(t)) + \frac{a_j(\tau)}{h(\tau)} S_* \left(e^k - e^{x}\right) \right].
\end{align}
The LHS of this equation is known due to the boundary conditions in \eqref{bc0} and \eqref{bc1}, hence \eqref{solInhomSup} finally takes the form
\begin{align} \label{solYB1}
 e^{k}\left[ 1 - a_j(t) \right] - \left[ 1 + \phi(t) - a_j(t)\right] e^{x_B(t)} = \frac{h(t)}{S_*} G(\tau(t), z(x_B(t)), x_B(t)).
\end{align}
Two equations \eqref{solInhomSup1} and \eqref{solYB1} should be solved together to determine the unknown functions $x_B(t)$ and $P_{xx}(t, x_B(t))$. Once this is done, the Put option price could be found by solving \eqref{solInhom}.

\subsubsection{Notes on solving \eqref{solInhomSup1}, \eqref{solYB1} and \eqref{solInhom}} \label{solvingSystem}

The system of equations in \eqref{solYB1}, \eqref{solInhomSup1} can be solved sequentially in the time $\tau \in [0, \tau(t)]$ (where usually $t=0$). Suppose we introduce a discrete grid in the time $\tau$ as $0, \tau_1, ... \tau_N = \tau(t)$ which contains $N+1$ nodes (in the simplest case the grid could be uniform). At every node $\tau_i, \ 0 \le i \le N$ the values of $x_B(\tau_i), P_{xx}(\tau_i, x_B(\tau_i))$ can be found iteratively in the following way. Starting with an initial guess for $P_{xx}(\tau_i), x_B(\tau_i)$, we substitute it into the RHS of \eqref{solInhomSup1}. Since at the time $\tau_i$ all values of $P_{xx}(\tau_i), x_B(\tau_k), \ 0 \le k < i$ are already known, \eqref{solInhomSup1} becomes a non-linear algebraic equation for $x_B(\tau_i)$. Solving it, we substitute thus found value of $x_B(\tau_i)$ into \eqref{solYB1} and obtain a linear Volterra integral equation of the second kind for $P_{xx}(\tau_i)$  which has been already discussed in above. Solving it, we obtain the next approximation for $P_{xx}(\tau_i)$ and continue the same way until convergence to a given tolerance is reached.

Once this is done, the Put option price can be found by solving \eqref{solInhom}. It is easy to see that the first two terms in the RHS of \eqref{solInhom} can now be computed explicitly. As far as for the third term is concerned, we can represent it in the form
\begin{align}
J(\tau,z) &= J(\tau^-,z) + J_1(\tau,x), \qquad \tau^- = \lim_{\epsilon \to 0^-} [\tau - \epsilon],
\end{align}
\noindent and having in mind that at $s = \tau$
\begin{equation}
\calG(\xi,\tau,z(\eta),\tau) = \delta(\xi - z) - \delta(\xi - 2 y(\tau) + z),
\end{equation}
\noindent finally obtain
\begin{align}
J_1 (\tau, x) &= \int_{x_B(\tau)}^{\infty} \frac{a_j(\tau) P(\tau,\xi)}{h(\tau)} \int_{x_B(\tau)}^{x} e^{\phi(\tau) \eta} \calG(\xi,\tau, z(\eta), \tau) d\eta d\xi \\
&=  \frac{a_j(\tau)}{h(\tau)} \int_{x_B(\tau)}^{x} d\eta \, e^{\phi(\tau) \eta}  \int_{x_B(\tau)}^{\infty} P(\tau,\xi) \left[\delta(\xi - \eta) - \delta(\xi - 2 x_B(\tau) + \eta ) \right] d\xi \nonumber \\
&=  - \frac{a_j(\tau)}{h(\tau)} \int_{x_B(\tau)}^{x} e^{\phi(\tau) \eta} P(\tau,\eta) d\eta \nonumber
\end{align}
This yields an alternative representation for \eqref{solInhom}
\begin{align} \label{solInhom2}
P(\tau,x) &+ a_j(\tau) e^{- \phi(\tau) x}  \int_{x_B(\tau)}^{x} e^{\phi(\tau) \eta} P(\tau,\eta) d\eta = e^{- \phi(\tau) x} \Bigg\{ S_*\left( e^{k} - e^{x_B(\tau)} \right) e^{\phi(\tau) x_B(t)} \\
&+ h(\tau) \int_{x_B(\tau)}^x e^{\phi(\tau) \eta} G(\tau, z(\eta), x_B(\tau)) d \eta
+ h(\tau) \int_0^{\tau^-} \int_{x_B(s)}^{\infty} \frac{a_j(s) P(s,\xi)}{h(s)} \calK_1(\tau, x, s, \xi) d\xi ds  \Bigg\} \nonumber \\
&\equiv e^{- \phi(\tau) x} \Theta(\tau, x). \nonumber
\end{align}

But the term $J(\tau^-,z)$ is already known at time $\tau$\footnote{If one works on a discrete grid and $\tau = \tau_i$, then $\tau^-$ is simply $\tau_{i-1}$} and, hence, $\Theta(\tau, x)$ is just a source term. Therefore, \eqref{solInhom} is an ODE with respect to $P(\tau,x)$ that can be solved analytically to yield
\begin{align} \label{PsolExp}
P(\tau,x) &= e^{- (a_j(\tau) +\phi )x} \Bigg[
P(\tau,x_B(\tau)) e^{(a_j(\tau) + \phi(\tau)) x_B(\tau)} + e^{a_j(\tau) x} \Theta(\tau, x) - e^{a_j(\tau) x_B(\tau)} \Theta(\tau, x_B(\tau)) \\
&- a_j(\tau) \int_{x_B(\tau)}^x e^{a \eta} \Theta(\tau, \eta) d\eta \Bigg], \qquad P(\tau,x_B(\tau)) = S_*\left( e^{k} - e^{x_B(\tau)}\right). \nonumber
\end{align}

Thus, once \eqref{solInhomSup1}, \eqref{solYB1} are solved and $x_B(\tau), P_{xx}(\tau, x_B(\tau))$ are found, no other steps should be taken since \eqref{solInhom} can be solved analytically, and the American Put option price is represented by \eqref{PsolExp}.

\section{The Kou model} \label{secKou}

The Kou model, proposed in \citep{Kou2004}, is a double exponential jump model with the \LY density
\begin{equation} \label{Kou}
\nu (dy) = \lambda\left[ p \theta_1 e^{-\theta_1 y} {\bf 1}_{y \ge 0} + (1-p) \theta_2 e^{\theta_2 y} {\bf 1}_{y < 0} \right] dy,
\end{equation}
\noindent where $\theta_1 > 1$, $\theta_2 > 0$, $1 > p > 0$, $\lambda \ge 0$. The first condition is imposed to ensure that the stock price $S_t$ has finite expectation.

Here we wish to show that the approach proposed in Section~\ref{ExpJumps} can also be utilized to price American options written on the stock which follows a jump-diffusion model where the jump part is represented by the Kou model. The difference with the exponential model is obvious: the Kou model allows both negative exponential jumps (occurred with the probability $p$) and positive exponential jumps (occurred with the probability $1-p$), while the pure exponential model supports only one -sided jumps (either negative or positive).

Substituting the density in \eqref{Kou} into the definition of the operator $\mathcal{J}$ in \eqref{intGen} and carrying out the integration (recalling that we treat $\partial/ \partial x$ as a constant) yields
\begin{align} \label{KouJ}
\mathcal{J} &= \lambda  \left[ -1 + \mu \triangledown_x + p \theta_1 (\theta_1 - \triangledown_x)^{-1}
+ (1-p) \theta_2  (\triangledown_x + \theta_2)^{-1}\right],\qquad
\mu = \frac{p}{\theta_1 -1} + \frac{1-p}{\theta_2 + 1}.
\end{align}
An existence condition for $\mathcal{J}$ to be well-defined (i.e., for the integral in \eqref{intGen} to be well-defined
when using this \LY measure)  is $-\theta_2 <  Re(\triangledown_x) < \theta_1$, \citep{ItkinBook}. It could be treated, e.g., as follows: there exists a discrete approximation of $\triangledown_x$ such that all eigenvalues of matrix $A$ representing this operator in this discrete basis, obey this condition.

In this paper we extend this model by assuming the jump parameters $\theta_1, \theta_2, \lambda, p$ to be time-dependent for the same reason as it was explained in Section~\ref{ExpJumps}. Our focus here is on pricing an American Put option by using the time-dependent Geometric Brownian motion process for the diffusion part of $S_t$ and the time-dependent Kou model for jumps.

Let us assume that both resolvents $(\theta_1 - \triangledown_x)^{-1}$, $(\triangledown_x + \theta_2)^{-1}$  exist and are analytic and bounded, and introduce a new dependent variable
\begin{equation} \label{uDef}
u(t, x) = [(\theta_1(t) - \triangledown_x) (\theta_2(t) + \triangledown_x) ] P(t,x) =
\theta_1(t) \theta_2(t) P(t,x) + [\theta_1(t) - \theta_2(t)] P_x(t,x) - P_{xx}(t,x).
\end{equation}
Then, combining \eqref{PIDE}, \eqref{KouJ}, instead of a PIDE as in \eqref{PIDE}  we obtain a PDE
\begin{align} \label{KouProblemFull}
\fp{u}{t}  &+ \Big[r(t) - q(t) - \frac{1}{2}\sigma^2(t) + \lambda(t) \mu(t) \Big] \fp{u}{x} + \frac{1}{2}\sigma^2(t) \sop{u}{x} - [r(t) + \lambda(t)] u  + \kappa(t) P(t,x) + \beta(t) P_x(t,x) = 0, \nonumber \\
\beta(t) &= \lambda(t) [ p(t) \theta_1(t) - (1-p(t)) \theta_2(t)] + \theta'_2(t) - \theta'_1(t), \qquad
\kappa(t) = \lambda(t) \theta_1(t) \theta_2(t) + [\theta_1(t) \theta_2(t)]'_t.
\end{align}
This is the same type of equation as \eqref{PDE}. Therefore, this problem can be solved in the same way as described in Section~\ref{ExpJumps}, namely by again, using an extended version of the Duhamel's principle, which is described in Appendix~\ref{secDuhamel}. As \eqref{PDE} is written in terms of the function $u(t,x)$, the terminal and boundary conditions for $u(t,x)$ can be obtained from \eqref{tcFin}, \eqref{bc0} to yield\footnote{For setting the below condition we need to know the corresponding values of the option price, Delta and Gamma at $t=T$.}
\begin{align}
u(T,x) &= S_* e^x  \left\{ \frac{1}{2}\left[ \theta_1(t) - \theta_2(t) - 1\right] + \frac{1}{\sigma(T)} \delta(x - k) \right\}, \\
u(t,0) &= 0, \qquad u(t, x_B(t)) = S_* e^{x_B(t)}\left[\theta_1(t) - \theta_2(t) - 1\right]. \nonumber
\end{align}

Based on the analysis given in Appendix~\ref{secDuhamel}, we can solve \eqref{KouProblemFull} similar to how this was done in Section~\ref{ExpJumps}. First, solution of a homogeneous version of \eqref{KouProblemFull} has been already provided. Second, based on \eqref{odeDuhGen}, the last integral in \eqref{solInhom} can now be written as
\begin{align} \label{kouSource}
h(t) \int_0^\tau & \int_{x_B(s)}^{\infty} \frac{\kappa(s) P(s,\xi) + \beta(s) P_\xi(s,\xi) }{h(s)} \calK_1(\tau, x, s, \xi) d\xi ds \\
&= h(t) \int_0^\tau \frac{1}{h(s)} \int_{x_B(s)}^{\infty} \left[ \kappa(s) \calK_1(\tau, x, s, \xi) - \beta(s) \fp{\calK_1(\tau, x, s, \xi)}{\xi} \right] P(s,\xi) d\xi ds, \nonumber
\end{align}
\noindent because $P(s, \infty) = 0$ and $\calK_1(\tau, x, s, x_B(\tau)) = 0$. The derivative of $(\calK_1)'_\xi$ can also be calculated in closed form, see Appendix~\ref{app2}. Since we have already described all steps necessary to solve this problem, the remaining part of getting the explicit solution is left to the reader. The only difference is that, by definition of $u(t,x)$ in \eqref{uDef}, the ODE in \eqref{ode} should be replaced with
\begin{align} \label{odeKou}
\theta_1(\tau) \theta_2(\tau) P(\tau,x) + [\theta_1(\tau) - \theta_2(\tau)] P_x(\tau, x) - P_{xx}(\tau, x) &= h(\tau) G(\tau, x + f(\tau), x_B(\tau) + f(\tau)),
\end{align}
\noindent where $G(\tau, x + f(\tau), x_B(\tau) + f(\tau))$ also includes the integral in \eqref{kouSource}, similar to \eqref{solInhom}. This ODE can be integrated in closed form to yield
\begin{align} \label{solHomKou}
P(\tau,x) &= e^{- \theta_2(\tau) x} \left[ e^{x_B(\tau) \theta_2(\tau)} P(\tau, x_B(\tau)) + \calB(x_B(\tau)) -  \calB(x) -  \frac{1}{\theta_1(\tau) + \theta_2(\tau)} \int_{x_B(\tau)}^x e^{k \theta_2(\tau)}f(k) dk \right], \nonumber \\
\calB(x) &= \frac{e^{[\theta_1(\tau) + \theta_2(\tau)] x}}{\theta_1(\tau) + \theta_2(\tau)} \int_\infty^x e^{- k \theta_1(\tau)} f(k) dk, \qquad f(x) = G(\tau, x + f(\tau), x_B(\tau) + f(\tau)).
\end{align}
The solution in \eqref{solHomKou} obeys the boundary conditions in \eqref{bc0}.

Again, we obtain a system of two nonlinear algebraic equations for $x_B(t)$ and $P_{xx}(\tau, y(\tau))$ by first, differentiating
\eqref{solHomKou} by $x$ and then substituting $x \to x_B(t)$, and second by differentiating \eqref{solHomKou} twice by $x$ and then substituting $x \to x_B(t)$. This system can be solved as this is described in Section~\ref{solvingSystem}. Once done, the Put option price can be found from \eqref{solHomKou} in the same way how this was done at the end of Section~\ref{solvingSystem}.

\section{Solving the system of obtained equations} \label{solNonEqs}

Our results obtained in the previous Sections, reveal that pricing of American options under the jump-diffusion exponential models results in solving one {\it linear integral} Fredholm-Volterra equation of the second kind \eqref{solInhomSup1} for the option Gamma $P_{xx}(\tau,x)$, and then one {\it nonlinear algebraic} equation \eqref{solYB1} for the exercise boundary $x_B(t)$.

This is due to the fact that we solve this problem sequentially in time, i.e. given the initial condition we make a step  in time from $\tau=0$ to $\tau = \Delta \tau$, solve these two equations iteratively, proceed to the next step in time $\tau = 2 \Delta \tau$, and so on. At every time step, we start with the initial guess for $P_{xx}(\tau, y(\tau))$, and solve the algebraic equation \eqref{solYB1} independently. Once this is done, thus found $x_B(\tau)$ is substituted into \eqref{solInhomSup1} which then becomes a {\it linear} Fredholm-Volterra equation for $P(\tau,x)$. It can be solved numerically to obtain a new value of $P_{xx}(\tau, y(\tau))$, etc. until convergence is achieved.

Compared with the approach in \citep{Chiarella2009} where the Merton jump-diffusion model with constant coefficients was considered, it is important to underline the advantages of our method: i) here all parameters of the model are not constant but time-dependent; ii) instead of solving a system of two nonlinear integral equations for $P(\tau,x)$ and $x_B(\tau)$ here we need to solve a single nonlinear algebraic equation for $x_B(\tau)$ and a linear integral equation for $P(\tau,x)$.  Besides, at every step in time we don't need to compute the RHS integrals from scratch since we use its value from the previous step and simply add an extra term with $s = \tau$.

Also, a similar set of integral equations was obtained in \citep{ItkinMuravey2023} for the time dependent stochastic Verhulst model, therefore, all the advantages of our approach remain to be valid in this case as well. A survey of various methods on how this type of the Volterra equations can be efficiently solved numerically, see \citep{ItkinMuravey2023}.

In the same way the American Call option price can be found but now considering a spatial domain $x \in (-\infty, x_B(t)]$. For the detailed explanation, again see \citep{ItkinMuravey2023}.

\subsection{Numerical method of solving \eqref{solYB1}, \eqref{solInhomSup1}} \label{numEx}

To illustrate our approach, we consider a jump-diffusion model with exponential jumps described in Section~\ref{ExpJumps} and numerically solve the system of equations in \eqref{solYB1}, \eqref{solInhomSup1} (again, see \citep{ItkinMuravey2023} for survey of various numerical methods that can be used for doing so). In particular, a collocation  method with some appropriate basis functions would be a natural choice to use. In this method the unknown option price $P(t,x)$ can be represented as a truncated series on basis functions with $N$ terms with yet unknown coefficients  $a_1(t),\ldots,a_N(t)$ to be functions of the time $t$, see survey in \citep{Amin2020} among others.  Then, a sequential in time procedure described in above can be applied.

\begin{figure}[!htb]
\begin{center}
\subfloat[]{\includegraphics[width=0.54\textwidth]{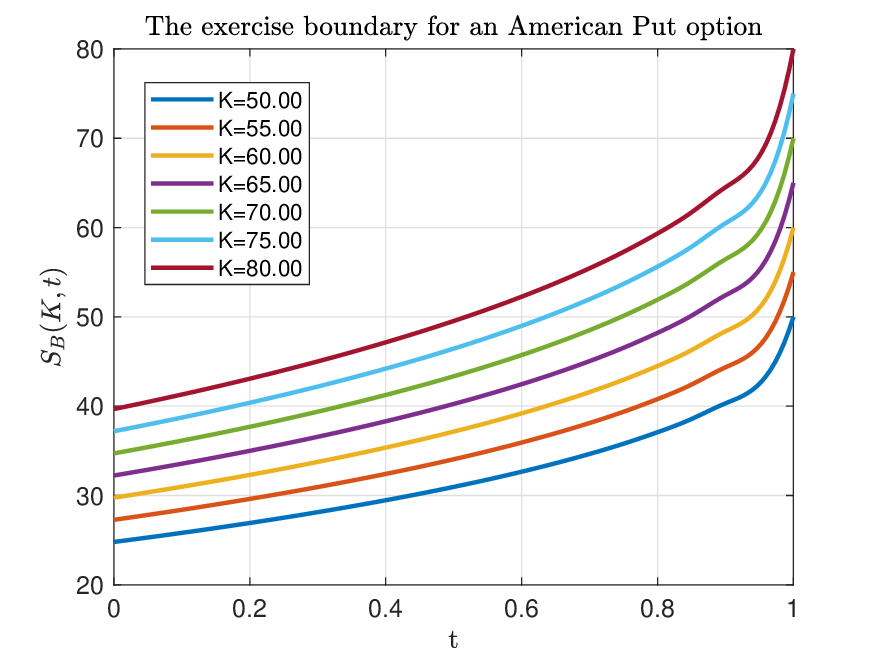}}
\hspace*{-0.3in}
\subfloat[]{\includegraphics[width=0.54\textwidth]{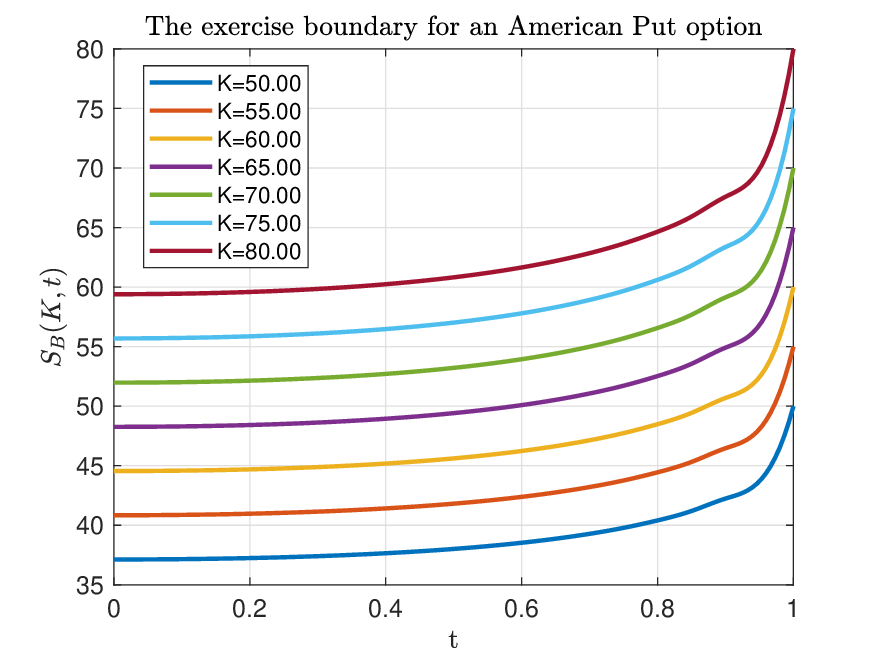}}
\end{center}
\vspace{-1em}
\caption{Exercise boundary $S_B(t)$ of the American Put option computed for various strikes $K \in [50:80]$; (a) parameters of the test are given in Table~\ref{tab1}, but no jumps; (b) same but with jumps.}
\label{test}
\end{figure}

Without any loss of generality, we choose the following dependencies of the model parameters on time
\begin{equation*} \label{ex}
r(t) = r_0 e^{- r_k t}, \quad q(t) = q_0, \quad \sigma(t) = \sigma_0 e^{-\sigma_k t}, \quad
\lambda(t) = \lambda_{0} + \lambda_{k} t, \quad \phi(t) = \phi_0 +  \phi_1 t^2.
\end{equation*}
Here $r_0, r_k, q_0, \sigma_0, \sigma_k, \lambda_0, \lambda_k, \phi_0, \phi_k$ are constants given in Table~\ref{tab1}.
The \eqref{solYB1}, \eqref{solInhomSup1} are solved by using the trapezoid quadratures to approximate the integrals (certainly, higher order quadratures can be used here as well to improve accuracy of the method). We run the test for a set of strikes $K \in [50, 55, 60, 65, 70, 75, 80]$ and choose $N=12, L=10$, where $L$ is parameter of the exponential Legendre polynomials, \citep{tauMethod2016}. The integral on $\xi$  in \eqref{solInhom}, \eqref{solInhomSup1} is computed by using the trapezoid rule with the truncated upper limit at $x_B(\tau) + 4 L$ and 30 points in $\xi$.
 \begin{table}[!htb]
\begin{center}
\begin{tabular}{|c|c|c|c|c|c|c|c|c|c|c|c|c|}
\hline
$r_0$ & $r_k$ & $q_0$ &
$\sigma_0$ & $\sigma_k$ & $\lambda_0$ & $\lambda_k$ &
$\phi_0$ & $\phi_k$  & $T$   \\
\hline
0.03 & 0.01 & 0.02 & 0.5  & 0.2 & 0.4 & 0.01 & 0.2 & 0.1  & 1 \\
\hline
\end{tabular}
\vspace{1em}
\caption{Parameters of the test.}
\vspace{-1em}
\label{tab1}
\end{center}
\end{table}
In Fig.~\ref{test} the exercise boundaries for the American Put option computed in this experiment are presented as a function of the time $t$. For the initial guess of $x_B(t_i)$ the already computed value $x_B(t_{i-1})$ can be chosen, and then the method typically converges within 5-6 iterations. The total elapsed time to compute $S_B(t)$ on a temporal grid: $t_i  = i \Delta t, \ i \in [0,M], \ \Delta t = T/M$ for all strikes with $M=20$ is 0.07 sec  per strike in Matlab using two Intel Quad-Core i7-4790 CPUs, each 3.80 Ghz. Despite our Matlab code can be naturally vectorized, we didn't do it since in this example we used a simple method (which can be improved in many different ways).
 This elapsed time is almost same as in the test shown in Fig.~\ref{EB} (also 0.07 sec per one strike, where the trinomial tree method was used to compute the exercise boundary under the Black-Scholes model with constant coefficients. However, for the trinomial tree the number of discrete stock values was equal to the number of discrete time values which is 400, otherwise the accuracy of calculations is insufficient.

Thus, the entire performance of the method is not worse than that of, e.g.,  the finite difference (FD) method applied to solving \eqref{PIDE} (especially, if the methods described in \citep{ItkinBook} are in use). However, the FD methods don't use an explicit form of the exercise boundary which is determined simultaneously with the American option price on a grid. In this case the American condition is checked as an extra step of the FD scheme, usually also iteratively (e.g., by using the "penalty" method, \citep{Halluin2004}), which also takes time. Finally, the typical accuracy of the  FD scheme is $O( (\Delta x)^2 + (\Delta t)^2)$ while solution of the integral equations can be provided with higher accuracy with no degradation of performance. Indeed, getting higher accuracy is one of our ultimate goals with this approach as that has been already claimed in Introduction.

 \section{Discussion}

In this paper we propose a semi-analytic approach to pricing American options for some time-dependent jump-diffusions models. The idea of the method is to further generalize our approach developed for pricing \textit{barrier} \citep{ItkinLiptonMuraveyBook}, and \textit{American} \citep{CarrItkin2020jd, ItkinMuravey2023}, options in various time-dependent one factor and even stochastic volatility models.

Aside from the academic interest, we strongly believe that the proposed approach also delivers various benefits for industrial applications when massive computation of American option prices has to be efficiently organized. We know that nowadays there exist a (to a certain extent) "naive" believe that modern numerical methods associated with finite-difference schemes or even with Monte-Carlo could serve for this purpose if they run at modern and fast computers or clusters. Unfortunately, this viewpoint doesn't take into account many practical aspects related to implementation of these schemes. As shown in detail in a recent presentation \citep{Andersen2023}, various problems associated with this approach cannot be solved uniformly and easily and require even more sophisticated and tricky algorithms and ideas.
The authors seriously ask questions like: a) Why are serious people still pricing American (equity) options using binomial trees? b) Why do standard-toolbox finite-difference methods perform so poorly? Can they be fixed
without excruciating pain? c) Can I find at least one order of magnitude improvement against existing methods? And improve Greeks? These questions and recommendation made in the presentation clearly demonstrate that construction of efficient numerical algorithms for pricing American options is still an actual problem.

 Perhaps, this should not be a big surprise since this type of approaches is not aware of the location of the exercise boundary and computes it implicitly as a part of the general algorithm. Since at the boundary the second derivative of the option price is discontinuous, this brings various numerical problems. Even the standard penalty method of \citep{Halluin2004} is relatively slow and reduces the accuracy of computations. In contrast, integral equations eliminate (or significantly eliminate) these problems while could be solved very fast and with high accuracy. Again, as per \citep{Andersen2023}, if possible (and it isn't always possible), one should use an integral method for American options, a la \citep{Andersen2016, AndersenLake2021}, because integral methods are spectrally converging and have numerous benefits in terms of speed, accuracy, avoidance of rounding errors, etc.

 However, despite the proposed approach looks promising, it is not clear at the moment how it can be extended for more sophisticated jump models, since it essentially relies on the linearity of the PDE obtained from the pricing PIDE by using a method of pseudo-differential operator, \citep{ItkinBook}.  To sum it all up, on the strong side of the approach are the following facts: i) it allows arbitrary dependencies of the model parameters on time; ii) it reduces solution of the pricing problem for American options to a simpler problem of solving an algebraic nonlinear equation for the exercise boundary and  a linear Fredholm-Volterra equation for the option price; iii) the options Greeks solve a similar Fredholm-Volterra linear equation  obtained by just differentiating \eqref{solInhom} by the required parameter.

On the weak side, this approach currently works only for some simple jump models. One way to circumvent this burden could be to relax time-dependence of the model parameters, i.e. treat them as constants.  In this case we should work with pseudo-parabolic equations of higher order with constant coefficients (or even with fractional pseudo-parabolic equations, e.g., for the CGMY model) that can be done by using various known integral transforms. However, the cost one has to pay for doing this is an expected puzzling with calibration of such a model to options market data, since most likely it won't be able to capture term-structure of the option implied volatility. Also, generalization of this approach, e.g., taking into account discrete dividends, especially for Call options could be complicated as well as efficient implementation will often require fairly sophisticated quadrature methods.

Furthermore, to underline, the integral equations and algebraic obtained in the paper, have not been known so far, and this is our contribution to the existing literature.

\section*{Acknowledgments}

I acknowledge a short discussion with Prof. Dilip Madan about possible usage of the GIT method for jump models, which induced the idea of this paper. Also, our long cooperation with Dmitry Muravey on development of the GIT method and multiple discussions are appreciated.

%%%%%%%%%%%%%%%%%%%%%%%%%%%%%%%%%%%%%%%%%%%%%%%%%%%%%%%%%%%%
%\printbibliography[title={References}]

%============== arxiv ============
%\bibliographystyle{apalike}
%\bibliography{tdOU}
%  Also commented out lines 212-290 in mydef2col.sty

\newcommand{\noopsort}[1]{} \newcommand{\printfirst}[2]{#1}
  \newcommand{\singleletter}[1]{#1} \newcommand{\switchargs}[2]{#2#1}

\appendixpage
\appendix
\numberwithin{equation}{section}
\setcounter{equation}{0}

\section{Calculation of the integrals in \eqref{solHom2}} \label{app3}

In this Appendix we derive an explicit representation for the functions $G(\tau(t), z(x_B(t)), x_B(t))$ and $G_z(\tau(t), z(x_B(t)), x_B(t))$ which appear in \eqref{solHom2}. As per \eqref{uFourier}, $G(\tau(t), z, x_B(t))$ is defined as follows
\begin{align} \label{Gdef}
G(\tau, z, y(\tau)) &\equiv I_1 + I_2 + I_3 = - S_* \frac{e^{- f(\tau)}}{2\sqrt{\pi \tau}}   \int_{y(0)}^{\infty} \gamma(\xi) e^{\xi} \left[e^{-\frac{( \xi - z)^2}{4\tau}} -  e^{-\frac{(\xi + z -2 y(\tau))^2}{4\tau}} \right] d\xi  \\
&-\int_0^\tau \frac{\Psi(s,y(s)) + y'(s) g(s)}{2\sqrt{\pi (\tau - s)}} \left[e^{-\frac{(z - y(s))^2}{4(\tau-s)}} -  e^{-\frac{(z -2 y(\tau) + y(s))^2}{4(\tau-s)}} \right] ds  \nonumber \\
&+ \int_0^\tau \frac{g(s)}{4\sqrt{\pi(\tau - s)^3}} \left[ (z - y(s)) e^{-\frac{(z - y(s))^2}{4(\tau -s)}} +
(y(s) + z - 2 y(\tau)) e^{-\frac{(z + y(s) - 2 y(\tau))^2}{4(\tau -s)}} \right] ds, \nonumber
\end{align}
\noindent and we compute the requested functions as a limit, i.e.,
\begin{equation*}
G(\tau(t), z(x_B(t)), x_B(t)) = \lim_{z \to y(\tau)^+} G(\tau, z, y(\tau)), \qquad \tau = \tau(t).
\end{equation*}

\subsection{$G(\tau(t), z(x_B(t)), x_B(t))$} \label{appG}

The result is already known from the boundary condition for the function $u(\tau, x_B(\tau))$. Nevertheless, here we verify it again by considering the corresponding limit $z \to y(\tau)$ in the definition of $G(\tau(t),z,x_B(t))$ in \eqref{uFourier}.

The first integral $I_1$ in \eqref{uFourier} doesn't have any singularity at $ z \to y(\tau)$ and, hence, vanishes at $z \to y(\tau)^+$. The third integral $I_3$ is a sum of two generalized heat potentials, see \citep{ItkinLiptonMuraveyBook} with $g(s)$ being the heat potential density. As shown in this book, section~8.1.1 assuming that $y'(\tau)$ is bounded, the result can be represented as
\begin{align} \label{HPint}
I_3 &= \frac{g(\tau)}{2} + \int_0^\tau g(k) \frac{y(\tau)- y(k)}{4\sqrt{\pi (\tau - k)^3}} e^{-\frac{(y(\tau)- y(k))^2}{4(\tau - k)}} dk \\
&+ \frac{g(\tau)}{2} + \int_0^\tau g(k) \frac{y(\tau)- 2y(\tau) + y(k)}{4\sqrt{\pi (\tau - k)^3}} e^{-\frac{(y(\tau)- 2y(\tau) - y(k))^2}{4(\tau - k)}} dk = g(\tau). \nonumber
\end{align}
As shown in \citep{TS1963}, the integral $\int_0^\tau \frac{y'(s) g(s)}{2\sqrt{\pi (\tau - s)}} e^{-\frac{(z - y(s))^2}{4(\tau-s)}} ds$ is continuous along the curve $z = y(\tau)$ because it converges uniformly and $y'(\tau)$ is bounded. Therefore, the term proportional to $y'(s) g(s)$ in $I_2$ vanishes. For the same reason, when $z \to y(\tau)^+$, the second derivative $P_{zz}(\tau,z)$ which appears in the definition of $\Psi(\tau, y(\tau))$ in \eqref{Psi}, is continuous in this area, and thus, the first term in $I_2$ also vanishes at $z \to y(\tau)^+$. Thus, we have
\begin{equation}
\lim_{z \to y(\tau)^+} G(\tau, z, y(\tau)) = g(\tau).
\end{equation}

\subsection{$G_z(\tau(t), z(x_B(t)), x_B(t))$} \label{appGz}

Differentiating \eqref{Gdef} by $z$ we obtain
\begin{align} \label{GdefZ}
G_z(\tau, & z, y(\tau)) \equiv  (I_1)_z + (I_2)_z + (I_3)_z \\
&= S_* \frac{e^{- f(\tau)}}{4\sqrt{\pi \tau^3}}   \int_{y(0)}^{\infty} \gamma(\xi) e^{\xi}  \left[  (z -\xi) e^{-\frac{(z-\xi)^2}{4\tau}}  - (z + \xi - 2 y(\tau) ) e^{-\frac{(z + \xi - 2 y(\tau))^2}{4\tau}} \right] d\xi  \nonumber \\
&+\int_0^\tau \frac{\Psi(s,y(s)) + y'(s) g(s)}{4\sqrt{\pi (\tau - s)^3}} \left[(z-y(s))e^{-\frac{(z - y(s))^2}{4(\tau-s)}} - (z -2 y(\tau) + y(s)) e^{-\frac{(z -2 y(\tau) + y(s))^2}{4(\tau-s)}} \right] ds  \nonumber \\
&- \int_0^\tau \frac{g(s)}{8\sqrt{\pi(\tau - s)^5}} \left[ (z - y(s))^2 e^{-\frac{(z - y(s))^2}{4(\tau -s)}} +
(z +y(s) - 2 y(\tau))^2 e^{-\frac{(z + y(s) - 2 y(\tau))^2}{4(\tau -s)}} \right] ds, \nonumber \\
&+ \int_0^\tau \frac{g(s)}{4\sqrt{\pi(\tau - s)^3}} \left[ e^{-\frac{(z - y(s))^2}{4(\tau -s)}} +
e^{-\frac{(z + y(s) - 2 y(\tau))^2}{4(\tau -s)}} \right] ds. \nonumber
\end{align}

The first integral is continuous at $ z \to y(\tau)$ and, therefore
\begin{equation}
\lim_{z \to y(\tau)} (I_1)_z = S_* \frac{e^{- f(\tau)}}{2\sqrt{\pi \tau^3}} \int_{y(0)}^{\infty} \gamma(\xi) e^{\xi}   (y(\tau) -\xi) e^{-\frac{(y(\tau)-\xi)^2}{4\tau}}d\xi.
\end{equation}

The second integral is a difference of two heat potentials, hence similar to  \eqref{HPint} we have
\begin{equation}
\lim_{z \to y(\tau)} (I_2)_z = \int_0^\tau \frac{\Psi(s,y(s)) + y'(s) g(s)}{2\sqrt{\pi (\tau - s)^3}} (y(\tau) -y(s))e^{-\frac{(y(\tau) - y(s))^2}{4(\tau-s)}} ds.
\end{equation}

The third integral converges uniformly, \citep{ItkinLiptonMuraveyBook}, and thus is a continuous function on the curve
$x = y(\tau)$. This implies that
\begin{equation}
\lim_{z \to y(\tau)} (I_3)_z = \int_0^\tau \frac{g(s)}{2\sqrt{\pi(\tau - s)^3}}
\left[ 1 - \frac{(y(\tau) - y(s))^2}{2 (\tau - s)} \right] e^{-\frac{(y(\tau) - y(s))^2}{4(\tau -s)}} ds.
\end{equation}

\subsection{The Volterra equation}

Combining the above representations and substituting them into \eqref{solHom2}, we obtain the integral Volterra equation of the second kind for $P_{xx}(\tau, x_B(\tau))$
\begin{align} \label{PxxVolterra}
P_{xx}(\tau,x_B(\tau)) &= S_* \phi^2(\tau)\left( e^{k} - e^{x_B(\tau)} \right) - h(\tau) g(\tau)
- h(\tau) \Bigg\{
S_* \frac{e^{- f(\tau)}}{2\sqrt{\pi \tau^3}} \int_{y(0)}^{\infty} \gamma(\xi) e^{\xi}   (y(\tau) -\xi) e^{-\frac{(y(\tau)-\xi)^2}{4\tau}}d\xi \nonumber \\
&+ \int_0^\tau \frac{e^{-\frac{(y(\tau) - y(s))^2}{4(\tau-s)}}}{2\sqrt{\pi (\tau - s)^3}}
\left[ \Psi(s,y(s)) (y(\tau) -y(s)) + g(s) \left[ 1 - \frac{(y(\tau) - y(s))^2}{2 (\tau - s)} + y'(\tau) \right] \right]
ds \Bigg\}, \nonumber
\end{align}
\noindent where $\Psi(\tau, y(\tau))$ is given by \eqref{Psi}
\begin{equation}
\Psi(\tau, y(\tau)) = \frac{1}{h(\tau)}\left[P_{zz}(\tau, y(\tau) - S_* \phi(\tau) e^{y(\tau) - f(\tau)}\right].
\end{equation}
This finally yields
\begin{align} \label{PxxVolterra1}
P_{xx}&(\tau,x_B(\tau)) = \calF(\tau,x_B(\tau)) - \int_0^\tau P_{xx}(s, y(s)) \frac{h(\tau) (y(\tau) -y(s)) }{2 h(s)} \frac{e^{-\frac{(y(\tau) - y(s))^2}{4(\tau-s)}}}{\sqrt{\pi (\tau - s)^3}} ds, \\
\calF&(\tau,x_B(\tau)) = S_* \phi^2(\tau)\left( e^{k} - e^{x_B(\tau)} \right) - h(\tau) \Bigg\{ g(\tau) +
S_* \frac{e^{- f(\tau)}}{2\sqrt{\pi \tau^3}} \int_{y(0)}^{\infty} \gamma(\xi) e^{\xi}   (y(\tau) -\xi) e^{-\frac{(y(\tau)-\xi)^2}{4\tau}}d\xi \nonumber \\
&- \int_0^\tau \frac{e^{-\frac{(y(\tau) - y(s))^2}{4(\tau-s)}}}{2\sqrt{\pi (\tau - s)^3}}
\Bigg[ \frac{S_*}{h(s)} \phi(s)  (y(\tau) -y(s)) e^{y(s) - f(s)} + g(s) \left[ 1 - \frac{(y(\tau) - y(s))^2}{2 (\tau - s)} + y'(\tau) \right] \Bigg]
ds \Bigg\}. \nonumber
\end{align}

\section{A wider view of the Duhamel's principle} \label{secDuhamel}

To remind, the basic Duhamel's formula provides solution of a non-homogeneous ODE
\begin{equation} \label{odeDuhamel}
u_t = A u + g, \qquad u(0) = f
\end{equation}
\noindent which may be expressed in terms of the solution operators of the corresponding homogeneous equation by the variation of parameters. This formula is given by the following theorem:

\begin{theorem}[\citep{evans10}]
 Suppose that $A: X \rightarrow X$ is a bounded linear operator on a Banach space $X$ and $\mathrm{T}(t)=e^{t A}$ is the associated uniformly continuous group. If $f \in X$ and $g \in C(\mathbb{R} ; X)$, then the solution $u \in C^1(\mathbb{R} ; X)$ of \eqref{odeDuhamel} is given by
\begin{equation} \label{odeSol}
u(t)=\mathrm{T}(t) f+\int_0^t \mathrm{~T}(t-s) g(s) ds.
\end{equation}
\end{theorem}

This solution is continuously strongly differentiable, satisfies \eqref{odeDuhamel} pointwise in $t$ for every $t \in \mathbb{R}$, and often is referred as a classical solution. For a strongly continuous group with an unbounded generator, however, the Duhamel formula \eqref{odeSol} need not define a function $u(t)$ that is differentiable at any time $t$ even if $g \in C(\mathbb{R} ; X)$.

This approach can be further generalized for semi-linear evolution equations of the type, \citep{Hunter2014}
\begin{equation} \label{ode2}
u_t = A u + g(u)
\end{equation}
\noindent where the linear operator $A$ generates a semigroup on a Banach space $X$ and
\begin{equation*}
g: \mathcal{D}(F) \subset X \rightarrow X
\end{equation*}
\noindent is a nonlinear function. For semi-linear PDEs, $g(u)$ typically depends on $u$ but none of its spatial derivatives and then (5.31) consists of a linear PDE perturbed by a zeroth-order nonlinear term.

If $\{\mathrm{T}(t)\}$ is the semigroup generated by $A$, we may replace \eqref{ode2} by an integral equation for $u:[0, T] \rightarrow X$
\begin{equation} \label{odeSol2}
u(t)=\mathrm{T}(t) u(0)+\int_0^t \mathrm{~T}(t-s) g(u(s)) ds.
\end{equation}
If the solutions of this integral equation exist and have sufficient regularity, then they also satisfy \eqref{ode2}.

The advantage of \eqref{odeSol} over \eqref{ode2} is that the unbounded operator $A$ is replaced by the bounded solution operators $\{\mathrm{T}(t)\}$. Moreover, since $\{\mathrm{T}(t-s)\}$  acts on $g(u(s))$, it is possible for the regularizing properties of the linear operators $\mathrm{T}$ to compensate for the destabilizing effects of the nonlinearity, see \citep{Hunter2014} for more examples.

The last but important point to mention is about the source term in \eqref{KouProblemFull} which is {\it linear} in  $P(t,x)$ and $P_x(t,x)$, i.e. we have $g(t, x, u, u_x) = \gamma(t) u + \beta(t) u_x$.  Under this circumstances the Duhamel's formula \eqref{odeSol2} is still valid. In more detail, see \citep{Penent2022} among others. The result reads
\begin{equation} \label{odeDuhGen}
u(t)=\mathrm{T}(t) u(0)+\int_0^t \mathrm{~T}(t-s) [\alpha(s) u(s) + \beta(s) u_s(s)] ds.
\end{equation}

\section{Solving a linear Fredholm-Volterra integral equation}

In this Appendix we shortly describe how \eqref{solInhomSup1} can be solved by using an appropriate collocation method which should be chosen by taking into account two important points: i) the spatial integral in \eqref{solInhomSup1} are taken over a semi-infinite interval; ii) the kernels in these integrals are exponential. Therefore, from this prospective we use the exponential Legendre Tau Method (ELTM) , see \citep{tauMethod2016} among others.

\subsection{Calculating spatial integrals in \eqref{solYB1}, \eqref{solInhomSup1} in closed form} \label{app2}

In this Section which is more technical, we want to show how the spatial integrals in \eqref{solYB1}, \eqref{solInhomSup1} can be computed in closed form. To recall, \eqref{solInhomSup1}, \eqref{solYB1},  read
\begin{align} \label{a1}
P(t,x) &= e^{- \phi(t) x} \Bigg\{ S_*\left( e^{k} - e^{x_B(t)} \right) e^{\phi(t) x_B(t)} + h(t) \int_{x_B(t)}^x e^{\phi(t) \eta} G(\tau(t), z(\eta), x_B(t)) d \eta  \\
&+ h(t) \int_0^\tau \int_{x_B(s)}^{\infty} \frac{a_j(s) }{h(s)}\calK_1(\tau, x, s, \xi) P(s,\xi) d\xi ds  \Bigg\}, \nonumber
\end{align}
\noindent and
\begin{align} \label{a2}
 e^{k}\left[ 1 - a_j(t) \right] - \left[ 1 + \phi(t) - a_j(t)\right] e^{x_B(t)} = \frac{h(t)}{S_*} G(\tau(t), z(x_B(t)), x_B(t)),
\end{align}
\noindent where
\begin{align} \label{gr1}
\calK_1(\tau, x, s, \xi) &\equiv \frac{1}{2 \sqrt{\pi (\tau - s)}} \int_{x_B(t)}^x e^{\phi(\tau) \eta} \left[e^{-\frac{(\xi - \eta - f(\tau-s))^2}{4(\tau-s)}} -  e^{-\frac{(\xi -2 y(\tau) + \eta + f(\tau-s))^2}{4(\tau-s)}} \right] d\eta \\
= \frac{1}{2} &e^{-\phi(\tau)[ \xi + f(\tau-s) - 2 y(\tau) - (\tau-s)\phi(\tau)]}
\left[ e^{2 \phi(\tau)(\xi - y(\tau))} A(x, \tau-s, \xi, \tau-s) -  A(x, \tau-s, 2y(\tau) - \xi, \tau-s) \right],  \nonumber \\
A(x, \nu, \xi, \omega) &= \erf \left(\frac{x + f(\nu) - \xi - 2\omega \phi (\tau)}{2 \sqrt{\omega}}\right)  - \erf\left(\frac{x_B(\tau) + f(\nu) - \xi - 2 \omega  \phi (\tau )}{2 \sqrt{\omega}}\right), \qquad \tau = \tau(t).  \nonumber
\end{align}

The first integral in \eqref{a1} can be further simplified to yield
\small{
\begin{align} \label{a3}
\int_{x_B(t)}^x & e^{\phi(t) \eta} G(\tau(t), z(\eta), x_B(t)) d \eta = \int_0^\tau \left[ \frac{g(s)}{4\sqrt{\pi(\tau - s)^3}} I_3  - \frac{\Psi(s,y(s)) + y'(s) g(s)}{2\sqrt{\pi (\tau - s)}} I_2 \right] ds
- S_* \frac{e^{- f(\tau)}}{4\sqrt{\pi \tau}}  I_1     , \nonumber \\
I_1 &= \int_{x_B(t)}^x e^{\phi(t) \eta} \int_{y(0)}^{\infty} e^{\xi}
\left[e^{-\frac{( \xi - \eta - f(\tau))^2}{4\tau}} -  e^{-\frac{(\xi + \eta + f(\tau) - 2 y(\tau))^2}{4\tau}} \right] d\xi d\eta \nonumber \\
&= B_1 + \sqrt{\pi \tau} \Big\{ \frac{e^{\tau +2 y(\tau) - f(\tau )}}{\phi(t) -1}  B_2 +
\frac{e^{\phi (t) (-f(\tau) - 2 \tau + y(0)) + y(0) - \tau}}{\phi(t) + 1}  B_3 \Big\}, \nonumber \\
B_1 &=  e^{\tau - f(\tau)} \left(\frac{e^{2 f(\tau)} \left(e^{x (\phi(t)+1)} - e^{x_B(t) (\phi(t)+1)}\right)}{\phi (t)+1}+\frac{e^{2 y(\tau )} \left(e^{x_B(t) (\phi (t)-1)}-e^{x (\phi (t)-1)}\right)}{\phi (t)-1}\right), \nonumber \\
B_2 &= - e^{(\phi (t)-1) (-f(\tau ) +\tau (\phi (t) + 1) + 2 y(\tau) - y(0))} A(x, \tau, -y(0) + 2 y(\tau), \tau) \nonumber \\
&+ e^{x (\phi (t)-1)} \erf\left(\frac{f(\tau) -2 \tau + x -2 y(\tau) + y(0)}{2 \sqrt{\tau}}\right) - e^{x_B(t) (\phi (t)-1)} \erf\left(\frac{f(\tau )-2 \tau + x_B(t) - 2 y(\tau ) + y(0)}{2 \sqrt{\tau }}\right), \nonumber \\
B_3 &= e^{x (\phi (t)-1)} \erf \left(\frac{f(\tau) -2 \tau + x-2 y(\tau )+y(0)}{2 \sqrt{\tau}}\right)
-e^{x_B(t)(\phi (t)-1)} \erf\left(\frac{f(\tau) -2 \tau + x_B(t) -2 y(\tau ) + y(0)}{2 \sqrt{\tau}}\right) \nonumber \\
&- e^{\tau  (\phi (t)+1)^2} A(x, \tau, y(0), \tau), \nonumber \\
I_2 &= \int_{x_B(t)}^x e^{\phi(t) \eta} \left[e^{-\frac{(\eta + f(s) - y(s))^2}{4(\tau-s)}} -  e^{-\frac{(\eta + f(s) - 2 y(\tau) + y(s))^2}{4(\tau-s)}} \right] d\eta = \sqrt{\pi (\tau -s)} e^{-\phi (t) (f(s) - (\tau - s) \phi(t) + y(s) - 2 y(\tau))} \nonumber \\
&\cdot \Big[e^{2 \phi(t) (y(s)-y(\tau))} A(x, s, y(s), \tau-s) - A(x, s, 2y(\tau) - y(s), \tau-s )\Big], \nonumber \\
I_3 &= \int_{x_B(t)}^x e^{\phi(t) \eta}\left[ (\eta + f(s) - y(s)) e^{-\frac{(\eta + f(s) - y(s))^2}{4(\tau -s)}} +
(y(s) + \eta + f(s) - 2 y(\tau)) e^{-\frac{(\eta + f(s) + y(s) - 2 y(\tau))^2}{4(\tau -s)}} \right] d\eta \nonumber \\
&= 2(\tau-s)\left\{ \phi(\tau) I_2 +
e^{\phi(\tau) x} \left[e^{-\frac{(x + f(s) - y(s))^2}{4(\tau-s)}} -  e^{-\frac{(x + f(s) - 2 y(\tau) + y(s))^2}{4(\tau-s)}} \right] \right\}. \nonumber
\end{align}
}%
\normalsize

One more integral that appears in the definition of $G(\tau, z, y(\tau)$, can be computed as follows
\begin{align}
- S_* \frac{e^{- f(\tau)}}{2\sqrt{\pi \tau}}   & \int_{y(0)}^{\infty} \gamma(\xi) e^{\xi} \left[e^{-\frac{( \xi - z)^2}{4\tau}} -  e^{-\frac{(\xi + z -2 y(\tau))^2}{4\tau}} \right] d\xi \\
&= \frac{1}{2} S_* e^{- f(\tau)} \left[ \erf \left(\frac{y(0) - z}{2\sqrt{\tau}}\right) - \erf \left(\frac{y(0) + z - 2y(\tau)}{2\sqrt{\tau}}\right)\right]. \nonumber
\end{align}

\subsection{Collocation method of solving a system of integral equations} \label{app2-2}

To solve \eqref{solInhomSup1} we use the collocation method, described in \citep{tauMethod2016}.
The main idea is in constructing functions approximation for the functional space $L_2([0,1])$ with the exponential Legendre basis. Then, every one- and two-variable functions are expanded in terms of the proposed basis functions.

The ELTM introduces a sequence of the exponential Legendre (EL) basis functions which are defined on the semi-infinite interval. In terms of our paper, we need to slightly modify this for the interval $[x_B(\tau), \infty)$ to get
\begin{equation}
E_n(x)=\calP_n \left(1-2 e^{-(x-x_B(\tau)) / L}\right),
\end{equation}
\noindent where $\calP_n(x)$ are the Legendre polynomials, \citep{as64}, and $L$ is a constant parameter which sets the length scale of the mapping. Functions $E_n(x)$ satisfy the following recurrence relation:
\begin{align} \label{recur}
E_0(x) &=1, \quad E_1(x)=1-2 e^{-(x-x_B(\tau)) / L}, \\
E_{n+1}(x) &= \left(\frac{2 n+1}{n+1}\right) \left(1-2 e^{-(x-x_B(\tau)) / L}\right) E_n(x)-\left(\frac{n}{n+1}\right) E_{n-1}(x), \quad n \geq 1. \nonumber
\end{align}

Exponential Legendre functions are orthogonal with respect to the weight function
\begin{equation*}
w_e(x) = \frac{2}{L} e^{-(x-x_B(\tau)) / L}
\end{equation*}
\noindent in the interval $[0,\infty)$, with the orthogonality property
\begin{equation}
\int_{x_B(\tau)}^{\infty} E_n(x) E_m(x) w_e(x) d x=\frac{2}{2 n+1} \delta_{n m},
\end{equation}
\noindent with $ \delta_{n m}$ being the Kronecker delta.

The classical Weierstrass theorem implies that the system $\{E_j(x)\}_{j \ge 0}$ is complete in the $L^2$ space. Thus, for any function $f(x) \in L^2$ the following expansion holds
\begin{equation} \label{series}
f(x) = \sum_{j=0}^{\infty} \alpha_j E_j(x), \qquad
\alpha_j = \frac{2 j+1}{2} \int_{x_B(\tau)}^{\infty} E_j(x) f(x) w(x) dx.
\end{equation}
If $f(x)$ in \eqref{series} is truncated up to the $N$ terms, then it can be written as
\begin{align} \label{serTrun}
f(x) &\simeq f_N(x) = \sum_{j=0}^{N-1} \alpha_j E_j(x) = A^\top E(x), \\
A &= [\alpha_0, \alpha_1, \ldots, \alpha_{N-1}]^\top, \qquad
E(x) = [E_0(x), E_1(x), \ldots, E_{N-1}(x)]^\top. \nonumber
\end{align}
To proceed, we substitute the representation \eqref{serTrun} into \eqref{a1} to obtain
\begin{align} \label{a1Ser}
\sum_{j=0}^{N-1} \alpha_j(t) E_j(x) &= e^{- \phi(t) x} \Bigg\{ S_*\left( e^{k} - e^{x_B(t)} \right) e^{\phi(t) x_B(t)} + h(t) \int_{x_B(t)}^x e^{\phi(t) \eta} G(\tau(t), z(\eta), x_B(t)) d \eta   \\
&+ h(t) \sum_{j=0}^{N-1}  \int_0^\tau \int_{x_B(s)}^{\infty}  \frac{\lambda(s) \phi(s) \alpha_j(s)}{h(s)} \calK_1(\tau, x, \xi, s) E_j(\xi)
d\xi ds  \Bigg\}, \nonumber
\end{align}

Similarly, substitution of \eqref{serTrun} into \eqref{a2} yields
\begin{align} \label{a2Ser}
\frac{S_*}{h(\tau)} \Big[\phi(\tau) \Big( e^{k} &- e^{x_B(\tau)} \Big) - \gamma(x_B(\tau)) e^{x_B(\tau)} \Big] =
G(\tau, z(x_B(\tau)), x_B(\tau)) \\
&+  \sum_{j=0}^{N-1}  \int_0^\tau \int_{x_B(s)}^{\infty} \frac{\lambda(s) \phi(s) \alpha_j(s)}{h(s)} \calK_2(\tau, x_B(s),s, \xi) E_j(\xi) d\xi ds, \qquad \calK_2(\tau, x_B(\tau),s, \xi) = \calG(\xi,\tau,s).  \nonumber
\end{align}
Here, $\calK_i(\tau, x, s, \xi), \ i=1,2$ are the integral kernels. Also, as it can be checked, the first integral in $G(\tau(t), z(x_B(t)), x_B(t))$ vanishes.

As this has been already explained in Section~\ref{solNonEqs}, all values of $\alpha_j(s), \ s < \tau$ in the RHS of \eqref{a2Ser}  are already known at the time $\tau$. At $s=\tau$, by the definition in \eqref{solInhomSup1}
$\calG(\xi, s, \tau)$ becomes the Dirac delta function, hence the integral on $\xi$ is gone, and instead in \eqref{a2Ser} we have
\begin{equation}
\frac{\lambda(\tau) \phi(\tau)}{h(\tau)} \sum_{j=0}^{N-1} \alpha_j(\tau) E_j(x_B(\tau)) =
\frac{\lambda(\tau) \phi(\tau)}{h(\tau)} S_*\left( e^{k} - e^{x_B(t)}\right).
\end{equation}
Thus, \eqref{a2Ser} becomes not an {\it integral}, but rather nonlinear {\it algebraic} equation with respect to $x_B(\tau)$ and can be solved independently.

Further on, thus found $x_B(\tau)$ can be substituted into \eqref{a1Ser}. By the same argument, $\calK_1(\tau, x, \tau, \xi)$ as defined in \eqref{gr1}, vanishes at $x = x_B(\tau)$ and at $x \ne x_B(\tau)$
\begin{equation*}
\calK_1(\tau, x, \tau, \xi) =
\begin{cases}
0 & x < \xi, \\
\frac{1}{2} e^{\phi(\tau) \xi} & x = \xi, \\
e^{\phi(\tau) \xi} & x > \xi.
\end{cases}
\end{equation*}
Accordingly,
\begin{equation*}
\int_{x_B(\tau)}^\infty \calK_2(\tau, x, \tau, \xi) E_j(\xi) d\xi =
\begin{cases}
0 & x < \xi, \\
\frac{1}{2}  \int_{x_B(\tau)}^\infty e^{\phi(\tau) \xi}  E_j(\xi) d\xi & x = \xi, \\
\int_{x_B(\tau)}^\infty e^{\phi(\tau) \xi}  E_j(\xi) d\xi & x > \xi.
\end{cases}
\end{equation*}

Therefore, by applying any quadrature rule to approximate the integral in $s$ in \eqref{a1Ser} and moving the terms with $s=\tau$ into the LHS, we obtain a linear system of equations for $\alpha_j(\tau), \ j=0,\ldots,N-1$.

\end{document}